# Assessment of FAIR (Findability, Accessibility, Interoperability, and Reusability) data implementation frameworks: a parametric approach

**Ranjeet Kumar Singh[1,2], Akanksha Nagpal[3], Arun Jadhav[3], Devika P. Madalli[4]**
[1]Delhi University Library System, University of Delhi, New Delhi- 110007
[2]DLIS , University of Calcutta, Kolkata, West Bengal- 700073
[3]CABI, NASC Complex, D.P. Shastri Marg, Pusa, New Delhi- 110012
[4]INFLIBNET Centre, Gandhinagar, Gujarat- 382421

**Abstract**
Open science movement has established reproducibility, transparency, and validation of research outputs as essential norms for conducting scientific research. It advocates for open access to research outputs, especially research data, to enable verification of published findings and its optimum reuse. The FAIR (Findable, Accessible, Interoperable, and Reusable) data principles, since its inception, support the philosophy of open science and have emerged as a foundational framework for making digital assets machine-actionable and enhancing their reusability and value in various domains, particularly in scientific research and data management. In response to the growing demand for making data FAIR, various FAIR implementation frameworks have been developed by various organizations to educate and make the scientific community more aware of FAIR and its principles and to make the adoption and implementation of FAIR easier. This paper provides a comprehensive review of the openly available FAIR implementation frameworks based on a parametric evaluation of these frameworks. The current work identifies 13 such frameworks and compares them against their coverage of the four foundational principles of FAIR, including an evaluation of these frameworks against a total of 36 parameters related to technical specifications, basic features, and FAIR implementation features and FAIR coverage. The study identifies that most of the frameworks only offer a step-by-step guide to FAIR implementation and are very technical in nature and seem to be adopting the technology-first approach, mostly guiding the deployment of various tools for FAIR implementation. Many frameworks are missing the critical aspects of explaining what, why, and how for the four foundational principles of FAIR, giving less consideration to the social aspects of FAIR. The study concludes that more such frameworks should be developed, considering the people-first approach rather than the technology-first.

## 1. Introduction

In a knowledge-based economy, ‘Knowledge is Power’ (Rodrígez García, 2001), and ‘Knowledge’ is the byproduct of ‘Information,’ which is ultimately derived from ‘Data,’ according to the well-known ‘DIKW paradigm’ (Peters et al., 2024). Therefore, claiming that "Data is power" in the context of the contemporary data-driven economy would not be a hyperbole of reality. Additionally, in today's linked world, data has become indispensable for routine operations and daily decision-making of people, enterprises, and organizations (including those engaged in governance and R&D). Further, the phrase "data is the new oil" (Humby, 2006), as coined by Clive Humby, is always used to describe how important data is to modern digital economies, just as oil was historical to industrialization and economic growth (Palmer, 2006; Stach, 2023). Furthermore, the same phrase is now transformed to “data is the new soil” (McCandless, 2010), as stated by David McCandless, where data is metaphorically visualized as a foundation for the industry, business, academia, and governance to bring new discoveries and opportunities, similar to soil working as the base for agriculture (Fraser, 2019; Shukla et al., 2022). Data is revolutionizing business, governance, and research by serving as the foundation for strategic planning, innovation, and decision-making across a wide range of sectors and is used as a base to maintain knowledge-based, information-rich, or data-driven economies (Singh & Madalli, 2023). Moreover, the rise of artificial intelligence (AI), machine learning, deep learning, generative AI, and other related fields, together with the revolution in data science, emphasize the significance of data and data management to a large degree.

In order to investigate intricate livelihood issues related to humanity and society, it is frequently imperative to get access to diverse data types and formats from open data repositories and archives. Public repositories confront enormous hurdles in ensuring that the scientific research community can readily find and use the data as the volume, variety, complexity, and veracity of (big) data increase (Singh, 2024). Consequently, the scientific community, about a decade ago, realized the importance of data, primarily research or scholarly data, and the critical aspects of open data sharing, (research) data management, and data stewardship to make (research) data available for reuse (Wilms et al., 2020; Zuiderwijk & Spiers, 2019) and the "Research Data management (RDM) has always been a challenge for research institutions" (Bellgard, 2020). It led to the stakeholders from various sectors like academia, R&D, businesses, funding bodies, publishing, etc., coming together to form the remarkable concept of "FAIR data principles," which are the four foundational principles—Findability, Accessibility, Interoperability, and Reusability—that serve to guide data producers, publishers, and managers as they navigate around various obstacles related to good data management practices (Wilkinson et al., 2016).

Under the aegis of open science, a worldwide ecosystem of actively findable, accessible, interoperable, and reusable digital objects, including data, services, and computing capability, is envisioned under the FAIR principles. The FAIR (Findability, Accessibility, Interoperability, and Reusability) principles provide guidelines ensuring that data and other digital objects, including software and codes, are accessible and usable at scale by humans as well as machines. The FAIR principles aim to enhance the ability to effectively find, access, exchange, and use the vast amount of data generated in research, governance, and businesses. Data is findable if one can quickly locate it, and it is accessible if one can get hold of it once it has been found. Further, data is interoperable if it is easily interpretable from a technical perspective, ideally by both humans and machines, and reusable defines what one can do with the data once it is accessed. Moreover, these principles strive to ensure that data is of high quality, properly managed, and effectively utilized across different domains.

There is an increasing need to harness scientific data effectively and implement FAIR principles efficiently, ensuring its optimal use and reuse. (Stall et al., 2019). The FAIR principles were explicitly designed for good practice around research data, emphasizing making research data machine-readable. Although, the 15 FAIR guiding principles aim to improve the findability, accessibility, interoperability, and reusability of digital assets rather than to mandate particular technology deployments. However, many researchers, scientists, and stakeholders from outside of the research community see implementing FAIR as a cumbersome task and regard this as highly technical (Betancort Cabrera et al., 2020; Jacob et al., 2021). To alleviate these problems and to facilitate the implementation of FAIR data principles, several FAIR implementation frameworks have been developed by numerous organizations associated with R&D (González et al., 2022). By supporting the provision of data that is Findable, Accessible, Interoperable, and Reusable, these principles overcome the limitations of classic data management strategies, which have proven insufficient concerning the ever-increasing volume and complexity of the data generated, leading to inefficiencies and missed research opportunities (Jacobsen et al., 2020, 2022). FAIR principles address various challenges related to data discovery, availability, integration, and reusability, often considered problematic in nearly all research domains. Creating rich metadata, making data openly accessible, and enabling data machine-readable and operable ensures that data sharing is seamless, eliminating frictional costs and maximizing the value of datasets. This structured approach to data management accelerates scientific research by providing high-quality, well-documented data, thus fostering more efficient and impactful research outcomes (Shanahan et al., 2021). These FAIR implementation frameworks help researchers from across domains with diverse degrees of expertise, knowledge, and background to quickly interpret the FAIR principles and implement them to meet their purpose of implementing FAIR.

### 1.1 Problem Statement

The complexities associated with the use or reuse of data, such as data in silos, inaccessibility of data, diverse data and metadata standards, domain-specific data standards, availability of various data file formats, diverse in-country data policies, and laws, etc., make it challenging for the data users to make the most out of data. Although Wilkinson et al. (2016) introduced the FAIR data principles to overcome these challenges, the original academic definition of FAIR and its principles are highly technical, with a particular emphasis on making data machine-readable and actionable. It led to other difficulties with "FAIR data implementation" as it is often perceived to be focusing only on technical aspects of making data FAIR. Numerous utility tools exist to do this already, and they usually work from the perspective of planning to get data ready for publication or designing a data management plan to make data FAIR. However, even to rationally use these tools, or where to use what, one needs a deep understanding of these tools along with the understanding of FAIR principles and its philosophies. Sometimes, the technicalities of these tools and the highly technical nature of FAIR principles and these tools become a discouraging factor for scholars coming from non-technical backgrounds, especially in Social Sciences, Management, Arts, and Humanities domains. Consequently, after the realization of the FAIR Data Principles, the scientific community, research groups, and various R&D organizations have developed various FAIR data implementation frameworks to serve various purposes related to FAIR data implementation, viz. FAIR Cookbook, FAIR4S, FAIR Digital Object Framework (FAIRDO), FAIR Decide Framework, FAIR Process Framework, etc. With so many frameworks available, selecting the right one for a specific application can be challenging and requires a robust comparative basis. This study aims to compare the popular FAIR implementation frameworks using a parametric approach, where frameworks are assessed based on specified parameters, to help users choose a framework based on their specific requirements. More specifically, the objectives of the current study are as follows:

- To identify the openly available FAIR implementation frameworks
- To identify the parameters that can be used to define and evaluate the FAIR implementation frameworks
- To explore the comprehensiveness and applicability of identified FAIR implementation frameworks based on their parameter coverage and present a comparison between these frameworks.
- To identify the relatedness between these tools based on selected parameters.

The remaining article is organized into the following sections: Section 2 provides a thorough review of the literature, divided into three parts: 2.1 FAIR and its Principles, 2.2 Approaches to FAIR Implementation, and 2.3 Related Works. This section identifies critical research gaps that frame the study. Section 3 details the methodology adopted for the research, offering a clear rationale for the chosen approach. Section 4 examines the parameters used to compare the identified FAIR implementation frameworks and presents the study's findings in detail. Section 5 explores the broader implications and relevance of the results, leading into Section 6, which concludes the article with key takeaways. A comprehensive list of references follows.

## 2. Literature Survey

### 2.1. FAIR and FAIR Principles

The paradigm of scientific research has shifted towards open science, and the goal of open science is to increase the reproducibility, transparency, reliability, visibility, accessibility, and upholding the integrity of scientific outcomes (Crüwell et al., 2019; Harper & Kim, 2018; Singh et al., 2022). Further, it advocates for making research findings, data, and other associated supplementary materials available to other peers and the general public for further discoveries, validation, and revalidation of research findings (Murray-Rust, 2008). The data revolution, which led to the data-driven economy, and the continuous demand to make research data openly available for reuse paved the path for the remarkable enunciation of FAIR data principles, which advocates that the data should be findable, accessible, interoperable, and reusable (Jacobsen et al., 2020; Wilkinson et al., 2016). Although open science necessitates open and free availability of data (Vicente-Saez & Martinez-Fuentes, 2018), however, applying the FAIR principles does not necessarily mean that data is shared openly or freely as FAIR and open data are not the same thing (Jati et al., 2022; Wagner et al., 2021; Wang & Savard, 2023); data can be FAIR or open or both or neither (Borgesius, Frederik Zuiderveen; Gray, Jonathan; Van Eechoud, Mireille, 2016). Whilst the most significant benefits may come from sharing FAIR and open data because this would support the widest reuse, there may be legitimate reasons or legal obligations for why data should not be shared openly (Borgesius, Frederik Zuiderveen; Gray, Jonathan; Van Eechoud, Mireille, 2016). Moreover, even when data cannot (or should not) be shared openly, the FAIR principles should still be utilized to maximize the possible reuse and utility of data by following the principle of 'as open as possible and as closed as necessary (Landi et al., 2020)'. Mostly, it can be inferred that open science mainly promotes more openness in scientific research (Allen & Mehler, 2019; Fecher & Friesike, 2014; Mirowski, 2018), whereas FAIR principles are meant to ensure the optimum reusability of data (Gajbe et al., 2021; Jacobsen et al., 2020; Singh & Madalli, 2023; Wilkinson et al., 2016).

The acronym 'FAIR' was first used in 2014 during a Lorentz workshop in the Netherlands (Jacobsen et al., 2020; Wilkinson et al., 2016); and the FAIR guiding principles were formally released in 2016 (Wilkinson et al., 2016). The FAIR principles came into place to enhance the utility of data by ensuring that it has been managed in a fashion that maximizes its usefulness and impact (Hasnain & Rebholz-Schuhmann, 2018; Wise et al., 2019). The core aspects of FAIR data indicate its importance in research studies (Devaraju et al., 2021). The first challenge that bursts forth while ensuring data reusability is finding the right data at the right time with light speed, i.e., findability of data, which means data must be easy to find (Řezník et al., 2022). Findability emphasizes the need for describing data using comprehensive metadata, assigning unique identifiers to datasets, and indexing them to some searchable repository to ensure data can be easily located (Ananthakrishnan et al., 2020; Pajouheshnia et al., 2024; Plomp, 2020; Wilkinson et al., 2016). Accessibility ensures that all data and resources are reachable using open protocols, which makes access smooth, including clear procedures for gaining authorization to access sensitive and restricted data (Landi et al., 2020; Shanahan & Bezuidenhout, 2022; Wilkinson et al., 2016). Interoperability ensures that data can be integrated and transferred across different systems and networks through formal and widely used languages for knowledge representation, standards, vocabularies, and accurate attributes (Lee et al., 2015; Wilkinson et al., 2016; Yang et al., 2020). Reusability requires that data be described in a way supporting its future use, including the terms of use by defining appropriate licenses, provenance information, ownership of data, and information regarding due credit if the data is reused (Jacobsen et al., 2020; Thanos, 2017; Wilkinson et al., 2016). Amongst all the four fundamental principles, interoperability is considered the most important one. If this is not achieved, the reusability cannot be ensured even though the dataset is easily findable and accessible (Kush et al., 2020). Implementing FAIR principles in research data management is very important, as it increases the impact of research by promoting the reuse of data, thus increasing the data visibility and citation rates, improving reproducibility, and encouraging collaboration (Gajbe et al., 2021; Singh et al., 2022; Singh & Madalli, 2023; Wise et al., 2019). These principles work together to foster efficiency, transparency, and collaboration within the research community, thereby increasing the progress in innovations and scientific discovery (Jacobsen et al., 2020).

### 2.2 FAIR Implementation

Since its inception, FAIR has become a buzzword among the scientific community, and the need to make data FAIR can be seen in the scientific literature across different domains (Van Vlijmen et al., 2020; Xing & Liu, 2022). To accomplish the notion of FAIR data, many organizations, viz. the European Union, Go FAIR Foundation, DeIC (Danish Infrastructure Cooperation), Pistoia Alliance, CABI, etc., came up with various initiatives and projects. These initiatives were spurred by the growing need and pressure among research stakeholders, especially those dealing with research data, to implement FAIR data principles in their research projects effectively. Consequently, it led to the development of various FAIR assessment tools—to assess how much FAIR a dataset is, such as FAIRdat, ARDC FAIR self-assessment, FAIRshake, F-UJI, etc.—, FAIRification tools—basically to help make data FAIR, such as FAIRScribe, FAIR-Aware, etc.—and FAIR implementation frameworks—mainly meant to educate scholars on FAIR and its principles and help them implement FAIR via a step-by-step guide, such as ACME-FAIR, FAIR hourglass, FAIR Cookbook, FAIR Process Framework, etc. Apart from this, a FAIR Data Maturity Model has been developed by one of the working groups of Research Data Alliance (RDA) to harmonize the FAIR assessment and is being adopted by various other organizations such as FAIRsFAIR to develop their metrics on FAIR data object assessment. Along with continuous studies related to rich metadata, standardized identifiers, open access protocols, and other features that further make it easier to retrieve and reuse data, all these recent developments and ongoing efforts are crucial for managing the increasing volume and complexity of research data and the need to ensure its compliance with FAIR principles.

The FAIR assessment tools make it easy to assess the FAIRness of any dataset and support the scientific community in verifying datasets for compliance with FAIR before their publication in a repository, which increases the discoverability and reusability of data (Bahim et al., 2020). Moreover, the FAIRification tools enhance data integration and interoperability through standardized vocabularies and formats that allow for the easy exchange of data across disciplines and maximize data reusability by ensuring well-documented data with clear licensing and provenance (Xu et al., 2023). In addition, the FAIR implementation frameworks are structured approaches that include guidelines, tools, and adherence to best practices to implement the FAIR principles in data management. They help researchers, institutions, and data custodians in the systematic implementation of FAIR principles, therefore ensuring that data is managed systematically, shared readily, and reused efficiently (Alharbi et al., 2023), including comprehensive guidelines on metadata standards, data licensing, and persistent identifiers, along with tools such as metadata registries and data repositories (Jacobsen et al., 2020). These frameworks also give guidance and assistance on evaluation metrics, such as the FAIR Data Maturity Model and FAIRmetrics, educational materials, including workshops and online courses, and encourage community participation through initiatives like GO FAIR Implementation Networks and the Research Data Alliance (RDA), which further enhance the holistic adoption and effectiveness of FAIR data approaches (Manola et al., 2019).

Many frameworks for the implementation of FAIR principles have been developed to ensure adherence to FAIR principles guiding data management. The core frameworks for implementing FAIR include FAIR-IMPACT, which systematically introduces FAIR principles at every phase or activity in research (Parland-von Essen & Dillo, 2022), FAIR hourglass model that broadens the general principles in the FAIR principles to specific actions (Schultes, 2023) and FAIR implementation Profile that is a customizable tool that provides a model for assessing the FAIR level of adherence (Schultes et al., 2020). Other frameworks include the FAIR Cookbook (Rocca-Serra et al., 2023) which acts as practical techniques for FAIRification, and FAIR4S (Shanahan et al., 2021) which is more detailed in describing FAIR practices while focusing on the European Open Science Cloud. Other frameworks, such as DeiC and FAIR Toolkit, provide guidelines and resources for FAIR for beginners and practical data management of the FAIR data life cycle, respectively (Principe et al., 2020). FOSTER provides an online course to learn about FAIR principles, while GO-FAIR helps organizations that prefer digital ways of having their FAIR way (Van Vlijmen et al., 2020). Other frameworks like the FAIR Implementation Choices and Challenges Model (Jacobsen et al., 2022; Mons et al., 2020) and FAIR Decide Framework (Alharbi et al., 2023) address decision-making to ensure that FAIR adoption is mainstreamed across institutions and health fields (Martínez-García et al., 2023). OpenAIRE and FAIR Digital Object Framework (FAIRDO) tools support institutions in making their data FAIR (Rettberg & Schmidt, 2012; Schultes, 2024). The recently launched FAIR process framework uses human-centered design and presents six steps to users that enable them to easily integrate FAIR principles into a project. Therefore, organizations and individuals need to adhere to these frameworks for data to eliminate the inconsistencies that impede data from reuse, ensuring open science and innovation are achieved.

Implementing FAIR principles increases data visibility and data citation, fosters collaboration, and helps researchers and institutions comply with funding agencies' mandates, ensuring high standards in data management practices (Shanahan et al., 2021).

### 2.3 Related Works

As time is ticking, the number of studies related to FAIR data—implementation, compliance, maturity, assessment, improvements, etc.—is increasing significantly, with a plethora of studies already available across the domains (Bahim et al., 2020). Especially in terms of studies related to the implementation of FAIR, there is a wealth of studies available discussing the development of FAIR implementation and assessment tools, guidelines, and other supporting materials. However, the literature suggests that significantly fewer efforts are being made to review or evaluate these tools, frameworks, and other materials.

Bahim et al. (2019) carried out an overall evaluation of 12 FAIR assessment tools ("the ANDS-NECTAR-RDS-FAIR data assessment tool, the DANS-Fairdat, the DANS-Fair enough?, the CSIRO 5-star Data Rating tool, the FAIR Metrics Questionnaire, the Stewardship Maturity Mix, the FAIR Evaluator, the Data Stewardship Wizard, the Checklist for Evaluation of Dataset Fitness for Use, the RDA-SHARC Evaluation, the WMO-Wide Stewardship Maturity Matrix for Climate Data, and the Data Use and Services Maturity Matrix") using 148 metrics that characterized these chosen tools. Further, Gehlen et al. (2022) expanded the study of Bahim et al. (2019) and based their study on five additional FAIR assessment tools ("Checklist for Evaluation of Dataset Fitness for Use, FAIR Maturity

Evaluation Service, FAIRshake, F-UJI, Self Assessment") from the research data repository's viewpoint and concluded that even while manual methods were more effective at gathering contextual information, there was no FAIR evaluation tool that could be used to evaluate discipline-specific FAIRness. Moreover, Krans et al. (2022) evaluated various FAIR assessment tools against their usability and performance. Their study consisted of 10 such tools, including "FAIRdat, FAIRenough?, ARDC FAIR self-assessment, FAIRshake, SATIFYD, Ammar, A. et al., FAIR evaluator software, RDA-SHARC, GARDIAN, Data Stewardship Wizard", and followed an expert evaluation method, employing three independent experts, against the defined assessment parameter to conclude the study. The study concluded that the development of various FAIR assessment tools gives flexibility to the scientific community to choose for their purpose, however these tools still need further improvements and defined use cases to gain better adaptability among users. In other studies, Bahlo (2024) and González et al. (2022) evaluated the AgReFed developed by the Agricultural Research Federation in Australia and FAIROs FAIR assessment tools, respectively. Their study explored various features of the tool, such as AgReFed provides FAIR matrics and FAIR scores for agricultural datasets through integrated automated F-UJI FAIR assessment API, and FAIROs provides a framework for assessing the compliance of a Research Object (and its constituents) against the FAIR principles via existing FAIR validators and proposes improved FAIR scoring methods for the FAIR assessment.

It is evident from the literature that although efforts have been made to evaluate various FAIR assessment tools and discuss the development of various FAIRification tools, significantly less or no effort has been made to evaluate the FAIR implementation frameworks developed to help, guide and educate the scientific community and ease the adoption and implementation of concepts related to FAIR principles. The current study aims to fill this void and systematically evaluate FAIR implementation frameworks against identified parameters. The current study is pervasive in its coverage and scope, presenting a novel approach to contribute to and advance the overall knowledge of FAIR implementation for the scientific community.

## 3. Methodology

### 3.1 Rationale

The current study is built upon a mixed approach, including descriptive and exploratory research methods. In order to produce an accurate and methodical description of specific instances, conditions, events, or the portrayal of what exists, descriptive research employs a number of methodologies, such as investigation via literature reviews, surveys, and so forth (Kothari & Garg, 2019; Singh et al., 2024). On the other hand, an exploratory study uses a range of methods, including exploration of literature, tools and techniques, methodologies, etc., to formulate research questions, generate ideas and concepts, and offer viable solutions to the problems it addresses. In the present study, the authors provide a cutting-edge report on various FAIR implementation frameworks primarily developed to simplify the FAIRification of data. The study identifies such frameworks from the literature and the web exploration along with parameters to evaluate these frameworks. It further provides a comparative evaluation of the frameworks by thoroughly exploring the identified frameworks against the identified parameters. Thus, given the nature of the problem undertaken in the study, adopting a mixed approach, including descriptive and exploratory research methods, which majorly involves literature searching, reviewing, identifying gaps, creating parameters for assessment, exploring relevant frameworks, gathering relevant information, and reporting, could be said to be the most suitable approach. The methodology is outlined in detail as follows:

### 3.2 Literature search strategy

An important component of finding research gaps is examining the literature, and every review procedure must carefully consider the search words to be used as "intuitively, the set of keyword terms defines a range of topics in which the user is interested (Doan et al., 2012)". We searched the databases Scopus, SpringerLink, Emerald Insight, Taylor & Francis Online, ACM digital library, EBSCOhost Research, and IEEE Xplore using the terms "FAIR", "Findability, Accessibility, Interoperability, Reusability", "Findable, Accessible, Interoperable, Reusable", "FAIR Implementation", "FAIR implementation framework", "FAIR Assessment", "FAIR Data", "FAIR Principles" both individually and in combination, using the Boolean operators, for the available literature on the topic. We restricted our search to the document type "article", language "English", and the period before "November 2024".

### 3.3 Literature identification

After deciding on the literature searching strategy, the actual search was performed using the advanced search option of the above-mentioned databases. Initially, the search retrieved a total of 17347 documents, which were further scrutinized using the advanced search option of the JabRef reference management tool for the terms "FAIR implementation", "FAIR assessment" "FAIR principles" and "FAIR data" (in the abstract field). A total of 737 documents were retrieved, and the titles and abstracts of the documents were manually read. Fifty-three relevant documents have been identified for the current study and included in the report.

### 3.4 Identification of parameters for assessment

There are various studies available on parametric assessment and approach. The authors selected a few of them, such as Gajbe et al. (2021), Bharti & Singh (2022), etc., to identify parameters for the assessment of FAIR implementation frameworks. The study identifies various parameters from the literature and framework's exploration, encompassing both technical and non-technical aspects crucial for effective FAIR implementation guidance and support. The study identified a total of 36 parameters, and these parameters were categorized into four major categories: technical parameters (16), basic features parameters (7), features related to FAIR implementation parameters (9), and FAIR coverage parameters (4). For technical aspects (see Table 1), the parameters include details such as the "first release" year, "latest release" year, and the "latest version" and its "status". Additional technical parameters involve the "regions covered" by the tool, the "license" under which it is registered, the level of "access" provided, whether it is "web-based," "source code availability," "offline accessibility," the available "form" (Document/Tool/Course), and its "architecture" (technical or socio-technical). These elements are significant as they demonstrate the framework's compliance with FAIR principles, its adaptability to different systems and user groups, its metadata support, security and privacy features, and its maintenance and open access status.

Table 1: List of Technical Parameters and their Definitions

| S. No. | Technical Parameters | All the details related to the technical aspects of the chosen frameworks |
|---|---|---|
| 1 | First release | Year, when the initial version was made available to the general public |
| 2 | Latest release | The most recent release year (up to November 2024). |
| 3 | Latest version | The current version of the tool available as of November 2024. |
| 4 | Status | Is the framework active/inactive |
| 5 | Regions covered | Specifies the geographical areas where the framework can be accessible and usable. |
| 6 | License | Indicates the licensing terms under which the tools or frameworks are registered. |
| 7 | Access | Defines whether the framework is open to the public or restricted to specific groups, such as researchers, staff, or students within a particular organization. |
| 8 | Web-based | Specifies whether the tools can be accessed and used through web browsers |
| 9 | Powered by/Infrastructure | The underlying technology or platform supporting the framework. |
| 10 | Technology stack | The set of technologies used to build and run the framework. |
| 11 | Source code availability | Determines whether the source code of the tools is available for download and installation on local machines. |
| 12 | Deployment options | The available methods for deploying the framework |
| 13 | Offline Accessibility | Whether the framework can be accessed offline |
| 14 | Form | Available in which form? Document/Tool/Course |
| 15 | Architecture | Technical or socio-technical |
| 16 | Audience | Prominent users of the framework or the targeted audience for whom the framework is built. |

Basic features parameters (see Table 2) related to the framework include “domain coverage” (whether domain-specific or independent), the availability of “documentation,” the presence of “community forums,” and the availability of help resources such as “FAQs/How Tos/User Guides/Support.” Furthermore, the framework's “feedback mechanism”, “case studies/success stories,” and “presence in scholarly literature" are essential for ensuring user support and broad adoption.

Table 2: List of Basic Features Parameters and their Definitions

| **S. No.** | **Parameters of Features: Basic** | **Basic features related to the framework** |
|---|---|---|
| 1 | Documentation | Availability of documentation of the framework |
| 2 | Community Forums | Does the framework have an active community forum? |
| 3 | FAQs/ How-Tos/ User Guide/ Support | Does the framework have a help section that offers FAQs/How-Tos/User Guide/ Support for various issues related to its use and technicalities? |
| 4 | Contribution options | Does the framework offer a mechanism that accepts feedback from the users and allows them to contribute to the framework? |
| 5 | Feedback Mechanism | Does the framework have a feedback mechanism to draw feedback from the users? |
| 6 | Case Studies/Success Stories | Available case studies or success stories |
| 7 | Presence in scholarly literature | The extent of availability or mention of the framework in scholarly literature from searched databases |

For FAIR implementation capabilities, parameters include (see Table 3) “defined steps” for implementation, guidance on identifying “blockers to FAIR” and “enablers to FAIR,” assistance in identifying “various data assets” that need to be made FAIR, and support for recognizing and integrating “in-country data policies.” The framework should also help develop “FAIR data governance strategies” and “FAIR technical implementation strategies,” offering “available courses/recipes” to aid users in achieving these goals.

Table 3: List of Parameters related to FAIR implementation and their definitions

| S. No. | Parameters of Features: FAIR Implementation | Features related to frameworks' FAIR implementation capability |
|---|---|---|
| 1 | Defined Steps for FAIR Implementation | Does the framework offer a defined step-by-step implementation plan or strategy? |
| 2 | Helps find Blockers to FAIR? | Does the framework guide the identification of various blockers for implementing FAIR? |
| 3 | Helps find Enablers to FAIR? | Does the framework guide the identification of various enablers for implementing FAIR? |
| 4 | Helps to Identify various data Assets? | Does the framework guide identifying various data assets of a project that need to be made FAIR? |
| 5 | Helps identify in-country data policies? | Does the framework guide identifying various in-country data policies for implementing FAIR? |
| 6 | Helps integrate in-country data policies? | Does the framework allow the integration of identified data policies for the robust implementation of FAIR? |
| 7 | Helps develop FAIR data governance strategies? | Does the framework assist in developing FAIR data governance guidelines, including data privacy, security, ethical and legal compliances, sensitivity and anonymization rules, etc., for the implementation of FAIR? |
| 8 | Helps develop FAIR Technical Implementation Strategies? | Does the framework assist in developing FAIR technical implementation plans by suggesting various technical options (tools and software) that are available? |
| 9 | Available Courses/Recipes | Does the framework offer any courses or recipes related to FAIR implementation? |

Finally, the parameters also assess the framework's guidance and support across all aspects of FAIR, including (see Table 4) "Findability," "Accessibility," "Interoperability," and "Reusability." Each aspect is evaluated based on what the principles entail, why they are important, how they can be implemented, and the available tools and technologies, including their pros and cons. This comprehensive evaluation ensures that the framework effectively supports data sharing, reusability, and research replicability while also fostering long-term sustainability, community adoption, and collaborative engagement within the research ecosystem.

Table 4: List of Parameters related to FAIR Coverage features and their Definitions

| S. No. | FAIR Coverage | Features related to frameworks' guidance or support on all aspects of FAIR and the extent of their coverage |
|---|---|---|
| 1 | Findability | Features related to frameworks' guidance or support on the Findability aspect of FAIR and the extent of its coverage |
| 1.1 | What | Explains what of all the principles related to Findability |
| 1.2 | Why | Explains why of all the principles related to Findability |
| 1.3 | How | Explains how of all the principles related to Findability |
| 1.4 | Tools | Suggest options of available tools and technologies with their pros and cons |
| 2 | Accessibility | Features related to frameworks' guidance or support on the Accessibility aspect of FAIR and the extent of its coverage |
| 2.1 | What | Explains what of all the principles related to Accessibility |
| 2.2 | Why | Explains why of all the principles related to Accessibility |
| 2.3 | How | Explains how of all the principles related to Accessibility |
| 2.4 | Tools | Suggest options of available tools and technologies with their pros and cons |
| 3 | Interoperability | Features related to frameworks' guidance or support on the Interoperability aspect of FAIR and the extent of its coverage |
| 3.1 | What | Explains what of all the principles related to Interoperability |
| 3.2 | Why | Explains why of all the principles related to Interoperability |
| 3.3 | How | Explains how of all the principles related to Interoperability |
| 3.4 | Tools | Suggest options of available tools and technologies with their pros and cons |
| 4 | Reusability | Features related to frameworks' guidance or support on the Reusability aspect of FAIR and the extent of its coverage |
| 4.1 | What | Explain what of all the principles related to Reusability |
| 4.2 | Why | Explains why of all the principles related to Reusability |
| 4.3 | How | Explains how of all the principles related to Reusability |
| 4.4 | Tools | Suggest options of available tools and technologies with their pros and cons |

### 3.5 Identification of FAIR implementation frameworks

The current study includes only openly available frameworks identified from the literature and web exploration. Also, it includes only those frameworks that are partially or fully built to support the implementation of FAIR. There are various kinds of materials available on FAIR implementation, such as study/learning materials, courses, guides, frameworks, etc., e.g., Foster. The authors have excluded all of them for this study and selected only those frameworks which are built with the intention of serving as an aid or assistant during the FAIRification of data, having well-structured step-by-step guidance on the implementation of FAIR. With these criteria, the study identified a total of 13 FAIR implementation frameworks, detailed below (see Table 5).

Table 5: Detailed list of identified FAIR implementation frameworks

| Code | Name of the Framework | Project & URL | Created By | Hosted By | Funded By | Partners | Domain | Language Coverage |
|---|---|---|---|---|---|---|---|---|
| F1 | FAIR implementation framework | FAIR-IMPACT (URL: https://www.fair-impact.eu/fair-implementation-framework) | European Open Science Cloud (EOSC) | European Open Science Cloud (EOSC) | European Union | DANS[1] | Agri-food, life sciences, photon & neutron science, social science and humanities | English |
| F2 | The FAIR Hourglass | NA (URL: https://content.iospress.com/articles/fair-connect/fc221514 ) | GO FAIR Foundation | GO FAIR Foundation | NA | NA | Domain Independent | English |
| F3 | Three-Point FAIRification Framework (M2M, FAIR Implementation Profile, FDP) | GO FAIR VODAN Implementation Network ENVRI-FAIR (URL: https://www.go-fair.org/how-to-go-fair/fair-impl | GO FAIR | GO FAIR | Go FAIR Foundation and European Union's Horizon 2020 research and innovation programme | NA | Domain Independent | English |

[1] **DANS**, DCC, SSI, DeiC, CSC, INRAe, Trust-IT Services, COMMpla, inria, SURF, Universität Bremen, CNRS, RDA, Observatoire de Paris, Universidad Politécnica de Madrid, DataCite, Aix-Marseille Université, Karlsruhe Institute of Technology, SikT, cessda, e-Science Data Factory, UK Data Service, LifeWatch ERIC, UK Research and Innovation, EMBL-EBI, Consiglio Nazionale delle Ricerche, CODATA, The University of Manchester, uefiscdi

| | | | | | | | | |
|---|---|---|---|---|---|---|---|---|
| | | ementation-profile/) | | | | | | |
| F4 | FAIR Cookbook | FAIRplus (URL: https://faircookbook.elixir-europe.org/content/recipes/introduction/fairification-process.html) | Collaboration [2] | ELIXIR | Innovative Medicines Initiative (IMI) | | Life Sciences | English |
| F5 | FAIR4S | EOSC FAIR4S Pilot Project (URL: https://eosc-fair4s.github.io/framework.html) | EOSC | EOSC & github.io | NA | NA | Domain Independent | English |
| F6 | FAIR for Beginners by DeiC | National FAIR Strategy (URL: https://www.deic.dk/en/data-management/instructions-and-guides/FAIR-for-Beginners) | DeIC | DeIC | NA | NA | Domain Independent | English & Dansk |
| F7 | ACME-FAIR | Fostering FAIR Data Practices in Europe (URL: https://www.fairsfair.eu/acme-fair-guide-rpo) | FAIRSFAIR | FAIRSFAIR | European Union's Horizon 2020 project | DANS[3] | Domain Independent | English |
| F8 | FAIRification Workflow | FAIR4Health (URL: https://fairtoolkit.pistoiaalliance.org/methods/fairification-workflow/) | NA | FAIR4Health | European Union's Horizon 2020 project | NA | Health | English |
| F9 | FAIR Toolkit | FAIR Implementation Project (URL: https://fairtoolkit.pistoiaalliance.org/practical-support-for-fair-data/) | Pistoia Alliance | Pistoia Alliance | AstraZeneca, novo nordisk, Bristol Myers Squibb, Pfizer, Roche | AstraZeneca[4] | Life Sciences | English |
| F10 | FAIR Implementation Choices and Challenges Model | NA (URL: https://osf.io/4v9pm/) | FAIR implementation choices and challenges (Erik Schultes, Tobias Kuhn, | Center for Open Science | NA | NA | Domain Independent | English |

[2] Collaboration with various academics, major (bio)pharmaceutical companies, and information and service companies partners

[3] DANS, CSC, DCC, STFC, EUA, SSI, DeiC, INRAe, Trust-IT Services, inria, SURF, Universität Bremen, CNRS, RDA, Universidad Carlos III de Madrid, DataCite, e-Science Data Factory, UK Data Archive, UK Research and Innovation, , CODATA, INES, DTL, European University Association, Georg-August Universität Göttingen, SPARC Europe, University of Helsinki, Universiteit Van Amsterdam, Universidade do Minho

[4] AstraZeneca, novo nordisk, Bristol Myers Squibb, Pfizer, Roche, Bayer, Elsevier, Lucidata, Curlew Research, epam, GO FAIR, SciBite, FAQIR, The Hyve, IQVIA, Newcastle University, nagarro, Novartis, ontoforce

| | | | | | | | | |
|---|---|---|---|---|---|---|---|---|
| | | | Annika Jacobsen.) | | | | | |
| F11 | FAIR Decide Framework | NA (URL: https://www.sciencedirect.com/science/article/pii/S1359644623000260) | Ebtisam Alharbi, a Ph.D. student at Manchester University | University of Manchester (Zenodo?) | Deanship of Scientific Research at Umm Al-Qura University, FAIRplus project, The Engineering and Physical Sciences Research Council (EPSRC) | NA | Pharmaceutical R&D Industry | English |
| F12 | FAIR Digital Object Framework | NA (URL: https://fairdigitalobjectframework.org/) | FAIRDO | Fair Digital Object Framework (FDOF) | FDO Forum | NA | Domain Independent | English |
| F13 | FAIR Process Framework | Enabling Data Access Project (URL: https://www.fairprocessframework.org/) | CABI | CABI | Bill & Melinda Gates Foundation | NA | Agriculture development | English |

### 3.6 Data reporting and analysis

The web pages of the identified frameworks (see Table 5) and the identified literature were explored to gather the value against each identified parameter. The data was recorded and organized using a spreadsheet for comparative analysis. It was subsequently divided into four separate tables based on the thematic parameters for further evaluation and reporting. These tables include descriptive, date, and categorical data. Descriptive data represents factual information; date data reflects the timeline; categorical data uses 'Y' to indicate known or available features and 'NA' to signify the absence of a feature or lack of available information on a specific parameter. The authors used various statistical analyses to find out interrelatedness among the frameworks based on their fulfillment of various parameters. The detailed findings are illustrated in the findings section of this study.

## 4. Findings

The current study adopts a parametric approach to assess various FAIR implementation frameworks and presents a comparative analysis between them. The current assessment is essential for all the stakeholders of data; viz., research organizations and researchers can identify their best framework as per their needs, whereas framework developers can use this study to reflect and introspect on their frameworks on what is missing or required to make it more comprehensive. The identified parameters have been categorized under four major themes, viz. technical specifications (see Table 1), basic features (see Table 2), features related to FAIR implementation (see Table 3), and FAIR coverage (see Table 4). The following subsections present the findings on these four thematic parametric assessments:

A total of 13 FAIR implementation frameworks have been identified for this study (see Table 5). It is observed that many of these frameworks are supported by significant projects, including FAIR-IMPACT and GO FAIR, and are funded by initiatives like the European Union’s Horizon 2020 program and donors like the Bill and Melinda Gates Foundation supporting the development of the FAIR Process Framework. It reflects a robust commitment to the advancement of FAIR principles within the scientific community. The frameworks have also been developed by a diverse range of creators and partners, from the European Open Science Cloud (EOSC) to major pharmaceutical companies such as AstraZeneca and Pfizer; these frameworks are products of extensive collaboration among academic, governmental, and industrial stakeholders. Partnerships with prestigious institutions further reinforce the global acceptance and integration of these frameworks. The hosting of these frameworks by reputable organizations like EOSC, ELIXIR, and the University of Manchester ensures their accessibility and reliability. More than half of the frameworks (7) are domain independent and can be applicable across domains, however, the FAIR implementation framework claims to be serving mostly the Agri Food, Life Sciences, Photon and Neutron Science, Social Science and Humanities domains, FAIR Cookbook and FAIR Toolkit is applied to the Life Sciences domain, FAIRification Workflow is made for the Health domain, FAIR Decide Framework is mostly suitable for the Pharmaceutical R&D Industry and FAIR Process Framework is for the Agriculture development sector.

**4.1 Technical Specifications**

The assessment of frameworks on technical parameters (Table 6) highlights the technical aspects of the identified frameworks against a total of 16 parameters. The release dates of these 13 identified frameworks span from 2019 to 2024, with FAIR4S and FAIR Implementation Choices and Challenges Model being the first two frameworks released in 2019 and FAIR Process Framework being the last framework introduced in 2024. The year 2020, four years after the inception of FAIR principles, became the most prominent year, bringing a total of six frameworks, including the Three Point FAIRification Framework (M2M, FAIR Implementation Profile, FDP), FAIR Cookbook, ACME-FAIR, FAIRification Workflow, FAIR Toolkit. The information about the latest release year for most of the frameworks is not available, however, most of the frameworks (12) are currently active and accessible to the users, indicating ongoing development and maintenance efforts. Some of these frameworks have multiple updates or versions, and information about their latest version is available for only three frameworks, including FAIR Cookbook, FAIR4S, and ACME-FAIR, exemplifying their ongoing efforts to keep these tools active and up to date, with recent updates ensuring their continued relevance.

All the identified frameworks assessed are designed for global use, offering open and often free access. Further, eight of the frameworks, viz. F3, F4, F5, F7, F9, F10, F11, and F13 are licensed under CC BY 4.0, and one of them (F8) is shared under CC BY-NC-ND 4.0 shows that these frameworks are meant for wide adaptability and applicability and to support researchers across various research communities and institutions. However, the accessibility of these frameworks is mainly web-based and is not available for offline access, emphasizing the importance of internet connectivity for their utilization. Moreover, it can be found that these frameworks are available in four main forms, viz. Web Page (F1, F3, F5, F6, F7, and F9), Working Document/Draft (F10 and F12), Conceptual Model (F2, F8, and F11), and Tool (F4, F11, and F13). More technically, these frameworks, especially those that are in the form of a tool, utilize diverse technologies and architectures. FAIR Cookbook uses the infrastructure of or is powered by Jupyter Book, FAIR4S by Workspace 7, ACME-FAIR by Zenedo, FAIR Implementation Choices and Challenges Model by Open Science Foundation (OSF), and FAIR Decide Framework by Qualtrics XM. The FAIR Cookbook uses varied technology stacks such as GitHub, Jupyter Book engine, Jupyter Notebooks, Markdown, HackMD, Binder, and Search Wizard for development and deployment of the tool, and it is made available for local deployment via Docker, and the source code is also made available through GitHub. The information regarding the technology stack, source code availability, and deployment options for the rest of the frameworks are unavailable. The FAIR4S and the FAIR Decide Framework follow a socio-technical architecture, whereas the FAIR Hourglass follows an hourglass architecture of the internet. Other than this, the FAIR implementation framework follows ACME-FAIR architecture, Three Point FAIRification Framework (M2M, FAIR Implementation Profile, FDP) follows a three-point framework architecture, along with FAIR for Beginners by DeiC and ACME-FAIR frameworks following the guides and instructions architecture. The FAIR Process Framework follows a people-first approach, focusing on downstream processes that are important for a project team to create a vision for making the data FAIR before moving into the technicalities of FAIR. Mostly, these frameworks are developed for a generic audience; however, some of the frameworks have specified their audiences as Principal Investigators, Data Managers, Data Scientists (FAIR Cookbook); Data Stewards, Laboratory Scientists, Business Analysts, Science Managers (FAIR Toolkit); research communities and Research institutions/organizations (FAIR4S and ACME-FAIR); pharmaceutical stakeholders and decision makers (FAIR Decide Framework); FAIR Process Framework aims to support the full potential of AgDev data, aligning with the Bill & Melinda Gates Foundation open access commitments to promote data availability, usability, and accessibility.

| **Technical Parameters** | **F1** | **F2** | **F3** | **F4** | **F5** | **F6** | **F7** | **F8** | **F9** | **F10** | **F11** | **F12** | **F13** |
|---|---|---|---|---|---|---|---|---|---|---|---|---|---|
| **First release** | 2023 | 2022 | 2020 | 2020 | NA | 2022 | 2020 | 2020 | 2020 | 2019 | 2023 | 2020 | 2024 |
| **Latest release** | NA | NA | NA | 2022 | 2019 | NA | 2022 | NA | NA | 2023 | NA | 2022 | 2024 |
| **Latest version** | NA | NA | NA | 0.1.0 | 2 | NA | 2.1 | NA | NA | NA | NA | NA | Beta |
| **Status** | A | A | A | A | NA | A | A | NA | A | A | A | A | A |
| **Regions covered** | Global | Global | Global | Global | Global | Global | Global | Global | Global | Global | Global | Global | Global |
| **License** | NA | NA | CC BY 4.0 | CC BY 4.0 | CC BY 4.0 | NA | CC BY 4.0 | CC BY-NC-ND 4.0 | CC BY 4.0 | CC BY 4.0 | CC BY 4.0 | NA | CC BY 4.0 |
| **Access** | O | O | O | O | O | O | O | O | Free | O | O | O | O |
| **Web-based** | Y | NA | Y | Y | Y | Y | Y | Y | Y | Y | Y | Y | Y |
| **Powered by/ Infrastructure** | NA | NA | NA | Jupiter Book | Workpackage 7 | NA | Zenedo | NA | NA | OSF-Open Science Framework | Qualtrics XM | NA | NA |
| **Technology Stack** | NA | NA | NA | GitHub, Jupyter Book engine, Jupyter Notebooks, Markdown, HackMD, Binder, Search Wizard | NA | NA | NA | NA | NA | NA | NA | NA | NA |
| **Source code availability** | NA | NA | NA | Yes (GitHub) | NA | NA | NA | NA | NA | NA | NA | NA | NA |
| **Deployment Options** | NA | NA | NA | Yes via Docker | NA | NA | NA | NA | NA | NA | NA | NA | NA |
| **Offline Accessibility** | NA | NA | NA | NA | NA | NA | NA | NA | NA | NA | NA | NA | Y |
| **Form** | Web Page | Conceptual Model (Research Paper) | Web Page | Tool | Web Page | Web Page | Web Page | Conceptual Model (Research Paper) | Web Page | Working Document | Conceptual Model & Tool | Working Draft/Document | Tool |
| **Architecture** | ACME-FAIR methodology | “hourglass” architecture of the Internet | Three Point Framework | NA | Socio-technical (Skills and competency framework) | Instructions and Guides | Guides | NA | NA | NA | Socio-technical | NA | Socio-technical |
| **Audience** | Generic | Generic | Generic | Principal Investigator, Data Manager, Data Scientist | Research communities and Research institutions | Generic | Research performing organizations | Generic | Data Stewards, Laboratory Scientists, Business Analysts, Science Managers | Generic | Pharmaceutical stakeholders and decision makers | Generic | Donors and project implementation teams in Agriculture development |

Table 6: Assessment of FAIR implementation frameworks against Technical Parameters

## 4.2 Basic features

The assessment of frameworks based on basic feature parameters (7 parameters) is done to evaluate their user-friendliness and the ability to attract community engagement (see Table 7). Most of the frameworks (F2, F4, F5, F6, F7, F9, F10, F11, F12, and F13) have documentation of the framework available whereas F1, F3, and F8 do not have the documentation in place. Only three frameworks (F4, F7, and F12) have a community forum for various discussions and queries related to the frameworks. In addition, FAQs/ How Tos/ User Guides/ Support is an essential feature from the user's point of view while adopting a framework and is provided by only six frameworks (F1, F4, F7, F9, F12, and F13). Further, the feedback mechanism, a way to communicate one’s feedback and reviews, is offered by six frameworks, including F1, F4, F7, F10, F12, and F13, and the contribution option is available only in five of the frameworks, which include F1, F4, F10, F12, and F13. All the 13 frameworks are available in the English language, with FAIR for Beginners by DeiC also available in Dansk language, highlighting a high level of adaptability of these frameworks in frequent English-proficient regions. Frameworks F1, F2, F3, F4, F6, F7, F9, and F13 also share the case studies and success stories related to the frameworks and frameworks F2, F3, F4, F8, and F11 also marked their presence in the scholarly literature (searched only in the Scopus Database), signifies their wide usability and applicability.

Table 7: Assessment of FAIR implementation frameworks against Basic Feature Parameters

| **Features: Basic** | **F1 (57%)** | **F2 (43%)** | **F3 (29%)** | **F4 (100%)** | **F5 (14%)** | **F6 (29%)** | **F7 (71%)** | **F8 (14%)** | **F9 (43%)** | **F10 (43%)** | **F11 (29%)** | **F12 (71%)** | **F13 (71%)** |
|---|---|---|---|---|---|---|---|---|---|---|---|---|---|
| **Documentation (77%)** | NA | Y | NA | Y | Y | Y | Y | NA | Y | Y | Y | Y | Y |
| **Community Forums (23%)** | NA | NA | NA | Y | NA | NA | Y | NA | NA | NA | NA | Y | NA |
| **FAQs/ How Tos/ User Guide/ Support (46%)** | Y | NA | NA | Y | NA | NA | Y | NA | Y | NA | NA | Y | Y |
| **Feedback Mechanism (46%)** | Y | NA | NA | Y | NA | NA | Y | NA | NA | Y | NA | Y | Y |
| **Contribution Options (38%)** | Y | NA | NA | Y | NA | NA | NA | NA | NA | Y | NA | Y | Y |
| **Case Studies/Success Stories (61%)** | Y | Y | Y | Y | NA | Y | Y | NA | Y | NA | NA | NA | Y |
| **Presence in scholarly literatures (Scopus) (38%)** | NA | Y | Y | Y | NA | NA | NA | Y | NA | NA | Y | NA | NA |

The table below (see Table 8) shows the similarity between the two frameworks based on these frameworks covering the basic feature parameters. The maximum similarity between the two frameworks based on basic feature parameters is seen to be 86% (between F2-F3, F2-F6, F2-F11, F3-F8, F5-F6, F5-F11, F6-F9, F8-F11, and F1-F13), and the minimum similarity was observed to be no similarity (0%) between F3 and F12. Further, F6 has more than 50% similarity with nine other frameworks (F2, F3, F5, F7-F11, F13), followed by F9 having a similarity of more than 50% with eight other frameworks, and F5, F11, and F13, having more than 50% similarity with seven other frameworks.

Table 8: Similarity between the frameworks based on coverage of basic features parameters

| | **F1** | **F2** | **F3** | **F4** | **F5** | **F6** | **F7** | **F8** | **F9** | **F10** | **F11** | **F12** | **F13** |
|---|---|---|---|---|---|---|---|---|---|---|---|---|---|
| **F1** | | 29% | 43% | 57% | 29% | 43% | 57% | 29% | 57% | 57% | 14% | 57% | 86% |
| **F2** | 29% | | 86% | 43% | 71% | 86% | 43% | 71% | 71% | 43% | 86% | 14% | 43% |
| **F3** | 43% | 86% | | 29% | 57% | 71% | 29% | 86% | 57% | 29% | 71% | 0% | 29% |
| **F4** | 57% | 43% | 29% | | 14% | 29% | 71% | 14% | 43% | 43% | 29% | 71% | 71% |
| **F5** | 29% | 71% | 57% | 14% | | 86% | 43% | 71% | 71% | 71% | 86% | 43% | 43% |
| **F6** | 43% | 86% | 71% | 29% | 86% | | 57% | 57% | 86% | 57% | 71% | 29% | 57% |
| **F7** | 57% | 43% | 29% | 71% | 43% | 57% | | 14% | 71% | 43% | 29% | 71% | 71% |
| **F8** | 29% | 71% | 86% | 14% | 71% | 57% | 14% | | 43% | 43% | 86% | 14% | 14% |
| **F9** | 57% | 71% | 57% | 43% | 71% | 86% | 71% | 43% | | 43% | 57% | 43% | 71% |
| **F10** | 57% | 43% | 29% | 43% | 71% | 57% | 43% | 43% | 43% | | 57% | 71% | 71% |
| **F11** | 14% | 86% | 71% | 29% | 86% | 71% | 29% | 86% | 57% | 57% | | 29% | 29% |
| **F12** | 57% | 14% | 0% | 71% | 43% | 29% | 71% | 14% | 43% | 71% | 29% | | 71% |
| **F13** | 86% | 43% | 29% | 71% | 43% | 57% | 71% | 14% | 71% | 71% | 29% | 71% | |

### 4.3 Features: FAIR Implementation

The third assessment of the frameworks was done based on the parameters (9 parameters) that signify features related to FAIR implementation. These parameters are basically questions or points of consideration that, if incorporated, can make the implementation of FAIR easy and smooth. Nine of the frameworks, including F1, F2, F3, F4, F5, F7, F8, F9, and F13, have a clearly defined step-by-by FAIR implementation stages. Identifying blockers or the pain points while implementing FAIR could be an essential precautionary measure, however, only five frameworks (F1, F4, F7, F9, and F13) help to identify the same. In addition, the identification of various enablers that can ease the implementation of FAIR could be a motivating factor for the users, but, again, only six frameworks (F1, F4, F5, F7, F9, and F13) help to find the same. Among all, si frameworks (F4, F5, F7, F8, F9, and F13) also assist in identifying the data assets, which is the primary component of FAIR implementation, and five of them (F4, F5, F7,F9 and F13) also guide to identify and integrate various data policies while implementing FAIR. Further, a few frameworks (F1, F4, F5, F7, F9 and F13) provide guidance in developing a FAIR data governance strategy, which is a great help for the sustainability of overall FAIR implementation. Furthermore, almost all the frameworks except the F11 help the users to develop FAIR technical implementation strategies, meaning they guide the users in choosing various technical options available for implementation against various FAIR principles. Moreover, five frameworks (F4, F6, F7, F9, and F13) also make available some courses or recipes to make their user understand the various principles of FAIR in a robust manner.

Table 9: Assessment of FAIR frameworks against features parameters related to FAIR implementation

| Features: FAIR Implementation | F1 (56%) | F2 (22%) | F3 (22%) | F4 (100 %) | F5 (78 %) | F6 (22%) | F7 (100%) | F8 (33%) | F9 (100%) | F10 (11%) | F11 (0%) | F12 (11%) | F13 (100%) |
|---|---|---|---|---|---|---|---|---|---|---|---|---|---|
| Defined Steps for FAIR Implementation? (69%) | Y | Y | Y | Y | Y | NA | Y | Y | Y | NA | NA | NA | Y |
| Helps find Blockers to FAIR? (38%) | Y | NA | NA | Y | NA | NA | Y | NA | Y | NA | NA | NA | Y |
| Helps find Enablers to FAIR? (46%) | Y | NA | NA | Y | Y | NA | Y | NA | Y | NA | NA | NA | Y |
| Helps to Identify various data assets? (46%) | NA | NA | NA | Y | Y | NA | Y | Y | Y | NA | NA | NA | Y |
| Helps identify various data policies? (38%) | NA | NA | NA | Y | Y | NA | Y | NA | Y | NA | NA | NA | Y |
| Helps integrate various data policies? (38%) | NA | NA | NA | Y | Y | NA | Y | NA | Y | NA | NA | NA | Y |
| Helps develop FAIR data governance strategies? (46%) | Y | NA | NA | Y | Y | NA | Y | NA | Y | NA | NA | NA | Y |
| Help develop FAIR Technical Implementation Strategies? (92%) | Y | Y | Y | Y | Y | Y | Y | Y | Y | Y | NA | Y | Y |
| Available Courses/Recipes? (38%) | NA | NA | NA | Y | NA | Y | Y | NA | Y | NA | NA | NA | Y |

The below table (see Table 10) shows the similarity between the two frameworks based on these frameworks covering the parameters related to features on FAIR Implementation. The maximum similarity between the two frameworks based on these parameters is seen to be 100% (between F2-F3, F4-F7, F4-F9, F4-13, F7-F9, F7-F13, F9-F13, and F10-F12) and the minimum similarity was observed to be no similarity (0%) between F4-F11, F7-F11, F9-F11, and F11-F13. Further, F1 has more than 50% similarity with ten other frameworks, followed by F8 having more than 50% similarity with eight other frameworks, whereas F2, F3, F10, and F12 have more than 50% similarity with seven other frameworks.

Table 10: Similarity between the frameworks based on coverage of features parameters related to FAIR implementation

| | F1 | F2 | F3 | F4 | F5 | F6 | F7 | F8 | F9 | F10 | F11 | F12 | F13 |
|---|---|---|---|---|---|---|---|---|---|---|---|---|---|
| **F1** | | 67% | 67% | 56% | 56% | 44% | 56% | 56% | 56% | 56% | 44% | 56% | 56% |
| **F2** | 67% | | 100% | 22% | 44% | 78% | 22% | 89% | 22% | 89% | 78% | 89% | 22% |
| **F3** | 67% | 100% | | 22% | 44% | 78% | 22% | 89% | 22% | 89% | 78% | 89% | 22% |
| **F4** | 56% | 22% | 22% | | 78% | 22% | 100% | 33% | 100% | 11% | 0% | 11% | 100% |
| **F5** | 56% | 44% | 44% | 78% | | 22% | 78% | 56% | 78% | 33% | 22% | 33% | 78% |
| **F6** | 44% | 78% | 78% | 22% | 22% | | 22% | 67% | 22% | 89% | 78% | 89% | 22% |
| **F7** | 56% | 22% | 22% | 100% | 78% | 22% | | 33% | 100% | 11% | 0% | 11% | 100% |
| **F8** | 56% | 89% | 89% | 33% | 56% | 67% | 33% | | 33% | 78% | 67% | 78% | 33% |
| **F9** | 56% | 22% | 22% | 100% | 78% | 22% | 100% | 33% | | 11% | 0% | 11% | 100% |
| **F10** | 56% | 89% | 89% | 11% | 33% | 89% | 11% | 78% | 11% | | 89% | 100% | 11% |
| **F11** | 44% | 78% | 78% | 0% | 22% | 78% | 0% | 67% | 0% | 89% | | 89% | 0% |
| **F12** | 56% | 89% | 89% | 11% | 33% | 89% | 11% | 78% | 11% | 100% | 89% | | 11% |
| **F13** | 56% | 22% | 22% | 100% | 78% | 22% | 100% | 33% | 100% | 11% | 0% | 11% | |

### 4.4 FAIR: Coverage

The fourth and final assessment of the frameworks was done based on their comprehensiveness in terms of covering various components of FAIR, i.e., Findability, Accessibility, Interoperability, and Reusability, and its principles (see Table 11). To assess the same, the authors explored the frameworks and determined whether or not the what, why, and how of the FAIR components had been adequately articulated, along with guidance about the adoption of various tools to implement various FAIR principles. It is evident from the table that frameworks F2, F3, F4, F6, F6, F10, and F13 cover all the what, why, how, and tools guidance aspects of the FAIR components. However, frameworks like F1, F5, F7, F8, and F11 do not cover the what, why, how, and tools guidance aspects of the FAIR components. Framework F12 does not cover the what, why, and how aspects but covers the tools guidance aspect of the FAIR components.

Table 11: Assessment of FAIR frameworks across FAIR coverage parameters

| FAIR Coverage | F1 | F2 | F3 | F4 | F5 | F6 | F7 | F8 | F9 | F10 | F11 | F12 | F13 |
|---|---|---|---|---|---|---|---|---|---|---|---|---|---|
| **Findability** | | | | | | | | | | | | | |
| **What** | NA | Y | Y | Y | NA | Y | NA | NA | Y | Y | NA | NA | Y |
| **Why** | NA | Y | Y | Y | NA | Y | NA | NA | Y | Y | NA | NA | Y |
| **How** | NA | Y | Y | Y | NA | Y | NA | NA | Y | Y | NA | NA | Y |
| **Tools** | NA | Y | Y | Y | NA | Y | NA | NA | Y | Y | NA | Y | Y |
| **Accessibility** | | | | | | | | | | | | | |
| **What** | NA | Y | Y | Y | NA | Y | NA | NA | Y | Y | NA | NA | Y |
| **Why** | NA | Y | Y | Y | NA | Y | NA | NA | Y | Y | NA | NA | Y |
| **How** | NA | Y | Y | Y | NA | Y | NA | NA | Y | Y | NA | NA | Y |
| **Tools** | NA | Y | Y | Y | NA | Y | NA | NA | Y | Y | NA | Y | Y |
| **Interoperability** | | | | | | | | | | | | | |
| **What** | NA | Y | Y | Y | NA | Y | Y | NA | Y | Y | NA | NA | Y |
| **Why** | NA | Y | Y | Y | NA | Y | Y | NA | Y | Y | NA | NA | Y |
| **How** | NA | Y | Y | Y | NA | Y | Y | NA | Y | Y | NA | NA | Y |
| **Tools** | NA | Y | Y | Y | NA | Y | Y | NA | Y | Y | NA | Y | Y |
| **Reusability** | | | | | | | | | | | | | |
| **What** | NA | Y | Y | Y | NA | Y | NA | NA | Y | Y | NA | NA | Y |
| **Why** | NA | Y | Y | Y | NA | Y | NA | NA | Y | Y | NA | NA | Y |
| **How** | NA | Y | Y | Y | NA | Y | NA | NA | Y | Y | NA | NA | Y |
| **Tools** | NA | Y | Y | Y | NA | Y | NA | NA | Y | Y | NA | Y | Y |

## 5. Discussion & Implications

Global spending on research and development (R&D) has reportedly surpassed $2.5 trillion annually. One of the crucial products of all R&D efforts is research data, which essentially acts as a foundation for scientific discoveries and findings as well as a critical resource for further studies. Unfortunately, a significant portion of the money spent on R&D activities is wasted due to insufficient data management strategies. It costs R&D organizations millions of dollars annually, which is effectively a significant financial loss for the global scientific and research community as well as a roadblock to scientific advancement. Inadequate data

management contributes to low-quality data by isolating data and preventing it from being found, accessed, or interoperable, resulting in very little to no reusability. In light of this, the idea of FAIR data and its guiding principles came to be seen as beneficial to the scientific community in terms of organizing research data and maximizing its potential by making it findable, accessible, interoperable, and reusable. Regardless of domain, the scientific community has been using these principles since their origin to include strong data management practices in their research endeavors. In addition, the necessity for data to be made accessible for future use across all disciplines gave rise to a number of FAIR implementation frameworks, which essentially act as guidelines for the scientific community in adhering to FAIR data standards.

The current study focused on the assessment of various FAIR implementation frameworks based on certain parameters identified. In order to address the first objective of the work, the study identifies 13 openly available and freely accessible FAIR implementation frameworks, listed in Table 5, from the web exploration and literature review. It is overwhelming to find out that these frameworks are developed as byproducts of various FAIR-oriented projects and a broad spectrum of R&D organizations from research to academia to industry, such as the European Commission, DANS, DataCite, Go FAIR Foundation, DCC, etc., are involved in the continuous development and improvement of these FAIR implementation frameworks. It is also observed that most of these frameworks are domain-independent (maybe adaptable for any domain) and are developed in the English language. Further, the study unveiled various parameters that can be used to evaluate or assess these frameworks, which is the second objective of this study. The study identified 38 parameters and categorized them into four major themes: technical specifications, basic features, FAIR implementation features, and FAIR Coverage (see Tables 1, 2, 3, and 4).

The main aim of the current study is to assess the identified FAIR implementation framework against the identified parameters. Most of the frameworks are in the form of a web page, do not have many technical backends, and are open and free to use with a CC BY 4.0 license. Apart from most of these frameworks being domain-independent, these frameworks cater to a wide array of disciplines, from life sciences and health to agri-food, photon and neutron sciences, and social sciences. Notably, frameworks such as FAIR-IMPACT cover multiple domains, providing robust documentation, user guides, FAQs, and case studies to support their adoption. Frameworks like the FAIR Cookbook and FAIR Toolkit, which are specifically focused on life sciences, stand out for their detailed documentation, extensive user support, and inclusion of success stories, illustrating practical applications of FAIR principles. The FAIR Process Framework provides a non-technical approach to FAIR and follows a people-first approach rather than focusing on the technological aspects. The FAIR Digital Object Framework (FDOF) and FAIR Decide Framework, which target domain-independent and pharmaceutical R&D industry users, respectively, also emphasize comprehensive documentation and support features. The frameworks such as FAIR for Beginners by DeiC offer valuable resources for new users.

Additionally, many frameworks provide feedback mechanisms and contribution options, which showcase their commitment to fostering a collaborative environment for continuous improvement. A critical aspect of these frameworks is their detailed guidance on FAIR implementation. Frameworks like FAIR-IMPACT, FAIR hourglass, FAIR Cookbook, and the FAIR process framework provide defined steps for FAIR implementation, helping users navigate the complexities of adopting these principles. Many frameworks also help identify blockers and enablers to FAIR implementation and offer strategic insights for overcoming challenges and leveraging opportunities. FAIR Cookbook, FAIR Toolkit, and FAIR process framework are noted for their comprehensive support in identifying and integrating various data policies and developing FAIR data governance and technical implementation strategies.

Case studies and success stories are included in several frameworks, such as the FAIR Cookbook and FAIR Toolkit and FAIR process framework, illustrating real-world applications and benefits of FAIR implementation. This practical evidence can be highly motivating for potential adopters. Moreover, the presence of several frameworks in scholarly literature, such as the FAIR hourglass, the FAIR Cookbook, etc., showcase their academic relevance and impact, demonstrating their value in advancing research and knowledge dissemination. Educational resources, including courses and recipes, are available in frameworks like the FAIR Cookbook and ACME-FAIR, providing users with practical learning opportunities to deepen their understanding of FAIR principles and their application.
In terms of findability, several frameworks like The FAIR hourglass, FAIR Process Framework, FAIR Cookbook, FAIR Toolkit, and FAIR process framework excel by providing comprehensive guidance on the what, why, and how of making data findable. These frameworks include detailed steps and tools that are necessary for implementing findability, such as metadata standards and persistent identifiers. The FAIR process framework and the FAIR Cookbook, in particular, stand out by offering clear instructions and practical tools to ensure data can be easily located and identified by users. This emphasis on findability is crucial for improving data visibility and facilitating efficient data discovery. Further, frameworks such as The FAIR process framework, The FAIR hourglass, and theFAIR Cookbook provide extensive resources to help users understand the importance of data accessibility, including the mechanisms to achieve it. They offer detailed guidelines and tools to ensure data can be accessed in a secure and reliable manner, thus promoting greater data sharing and collaboration. The availability of tools within these frameworks ensures that data can be accessed seamlessly while maintaining necessary controls and permissions. Furthermore, interoperability, which enables data to be integrated and used across different systems, is well-covered by frameworks like FAIR-IMPACT and The FAIR hourglass. These frameworks offer clear explanations and practical tools for achieving data interoperability, including standardized data formats and protocols. By facilitating interoperability, these frameworks ensure that data can be effectively combined and utilized in various research contexts, enhancing the potential for cross-disciplinary collaborations and innovations. Moreover, reusability is a fundamental aspect addressed by these frameworks to maximize the value of data. Frameworks such as FAIR4Health and FAIR Toolkit provide detailed guidance on ensuring data is reusable, including the necessary steps and tools to achieve this. These frameworks emphasize the importance of data quality, clear licensing, and documentation to ensure that data can be effectively reused by other researchers. The focus on reusability ensures that data can contribute to future research efforts, driving scientific advancements and efficiency. Above all, the recently launched FAIR Process Framework by CABI and the FAIR Cookbook covers various aspects of FAIR implementation in a very comprehensive manner and may be very easy and suitable to adapt by various research stakeholders across domains.
The final objective of this study is to measure the relatedness among these frameworks based on selected parameters. The relatedness among these parameters was measured based on the data captured in Table 8 and Table 10 using the concepts of XAND operators. To establish the relatedness or similarity between the two frameworks, the value against each parameter (whether the feature is available or not) in Table 7 and Table 9 is compared, respectively. Then after, the pairs of matching values (either Y & Y or NA and NA) for any two frameworks were counted and converted to their percentage equivalence (against the total number of parameters available in the Table). Further, heat map tables were created by plotting the calculated percentage similarity or relatedness between two frameworks (Table 8 and Table 10 for Table 7 and Table 9, respectively) for the representation purpose. It is observed from these heat maps that most of these frameworks share common features and are related or similar in many aspects.

The implications of this study are extensive and span across disciplines. Research efforts across disciplines produce valuable data as byproducts, and to maximize the reusability of this data, it is necessary to use FAIR principles or adopt effective data management methods. Therefore, the study is very novel in its nature and scope and it may be directly applicable and adaptable to all the researchers, scientists, data managers, general users of the data, information professionals, library professionals, etc. When choosing a FAIR implementation framework to include FAIR principles in their research projects, the scientific community in every discipline may find this study informative and interesting. Apart from this, the current study is also adaptable to organizations or research groups involved in FAIR implementation, especially to those who are involved in the development of similar frameworks like this. These organizations or research groups can have a retrospective view of their frameworks based on the findings of this study and they can adapt several new features, modify the frameworks, and publish an improved version of the framework. The current study would be a comprehensive resource for organizations wanting to develop a similar kind of FAIR implementation framework from scratch. Consequently, it may be urged that the current work has broad implications and would serve as a reference document to all FAIR data enthusiasts.

## Conclusion

The rapid rise of knowledge and resources regarding the FAIR data principles and their adoption by diverse stakeholders across disciplines is very encouraging for the advocates of Open Science and Open Data initiatives. Although these principles have been adopted widely, especially in research and academia, many stakeholders, particularly those involved in data generation, are still unaware of them (Oladipo et al., 2022). Hence, tailored training programs for data stewards are required to equip them with the necessary skills to apply the FAIR principles effectively. Recent developments in this area include the emergence of FAIR implementation frameworks for integrating FAIR in data-intensive research projects, ensuring successful adoption and application of the FAIR guiding principles. Although they vary in the extent of FAIR coverage, community engagement, and scholarly presence, these frameworks collectively highlight the global effort to advance FAIR principles, emphasizing open access, practical usability, and tailored support for diverse research communities. The open and often free access to these frameworks, coupled with their web-based accessibility, supports their broad applicability across different research communities and institutions. It may broaden the applicability of the frameworks to various users and promote a scientific ecosystem that is more inclusive and collaborative. Many organizations like the Go FAIR Foundation, CABI, European Open Science Cloud, Pistoia Alliance, FAIRSFAIR, DeIC, etc., and funders like the European Union, Bill & Melinda Gates Foundation, etc. are constantly engaged in various projects under the aegis of FAIR implementation in research projects across domains. Their engagement resulted in several FAIR implementation frameworks like the FAIR Cookbook, the FAIR Process Framework, the FAIR Toolkit, ACME-FAIR, FAIR4S, etc., helping the stakeholders in their journey towards FAIR adoption.

Above all, these frameworks should be strategically designed to implement the principles of Findability, Accessibility, Interoperability, and Reusability in a manner that emphasizes the needs and experiences of users. The starting point for FAIR data will need actors to learn about it in a variety of ways, and incentives for implementing FAIR data will change depending on the environment and the stakeholders. Hence, it is suggested that FAIR is context-dependent and can be viewed differently depending on the circumstances and context, and different countries, industries, domains, researchers, etc., may implement FAIR differently. Hence, in addition to the concentration on the technical components of adopting

FAIR, it is recommended that while developing any framework of this sort, it is essential to consider the non-technical or social aspects of FAIR implementation. Further, a human-centered design approach with people over technology would be a great addition to such initiatives in order to create an atmosphere that is conducive to the successful implementation of FAIR. With this strategy, scientists and researchers worldwide would be able to collaborate, participate in all research procedures, and share multiple significant scientific discoveries that benefit all aspects of human life.